\documentclass[aps,pra,onecolumn,superscriptaddress,nofootinbib]{revtex4}

\usepackage{graphicx}
\usepackage[normal]{subfigure}
\usepackage{caption}
\usepackage{latexsym}
\usepackage{amsmath}
\usepackage{amssymb}
\usepackage{amsfonts}
\usepackage{color}
\usepackage{pgf, tikz}
\usepackage{dsfont}
\usepackage{caption}

\usepackage[utf8]{inputenc}
\usepackage[english]{babel}

\begin{document}

\newcommand{\ket}[1]{\vert #1 \rangle}
\newcommand{\bra} [1] {\langle #1 \vert}
\newcommand{\braket}[2]{\langle #1 | #2 \rangle}
\newcommand{\proj}[1]{\ket{#1}\bra{#1}}
\newcommand{\mean}[1]{\langle #1 \rangle}
\newcommand{\opnorm}[1]{|\!|\!|#1|\!|\!|_2}
\newtheorem{theo}{Theorem}
\newtheorem{lem}{Lemma}
\newtheorem{defin}{Definition}
\newtheorem{corollary}{Corollary}
 \newtheorem{conj}{Conjecture}

\title{Entropy-power uncertainty relations : \\ towards a tight inequality for all Gaussian pure states}

\author{Anaelle Hertz}
\email{ahertz@ulb.ac.be}
\affiliation{Centre for Quantum Information and Communication, \'Ecole polytechnique de Bruxelles, CP 165, Universit\'e libre de Bruxelles, 1050 Brussels, Belgium}

\author{Michael G. Jabbour}
\affiliation{Centre for Quantum Information and Communication, \'Ecole polytechnique de Bruxelles, CP 165, Universit\'e libre de Bruxelles, 1050 Brussels, Belgium}

\author{Nicolas J. Cerf}
\affiliation{Centre for Quantum Information and Communication, \'Ecole polytechnique de Bruxelles, CP 165, Universit\'e libre de Bruxelles, 1050 Brussels, Belgium}

\begin{abstract}
We show that a proper expression of the uncertainty relation for a pair of canonically-conjugate continuous variables relies on entropy power, a standard notion in Shannon information theory for real-valued signals. The resulting entropy-power uncertainty relation is equivalent to the entropic formulation of the uncertainty relation due to Bialynicki-Birula and Mycielski, but can be further extended to rotated variables. Hence, based on a reasonable assumption, we give a partial proof of a tighter  form of the entropy-power uncertainty relation taking correlations into account and provide extensive numerical evidence of its validity.  Interestingly, it implies the generalized (rotation-invariant) Schrödinger-Robertson uncertainty relation exactly as the original entropy-power uncertainty relation implies Heisenberg relation. It is saturated for all Gaussian pure states, in contrast with hitherto known entropic formulations of the uncertainty principle. 
\end{abstract}

\maketitle



\section{Introduction}

The uncertainty principle lies at the heart of quantum physics. It exhibits one of the key divergences between a classical and a quantum system. Classically, it is in principle possible to specify the precise value of all measurable quantities simultaneously in a given state of a system. In contrast, whenever two quantum observables do not commute, it is impossible to define a quantum state for which their values are simultaneously specified with infinite precision. A paradigmatic example is given by Heisenberg's original formulation of the uncertainty principle expressed in terms of variances of two canonically-conjugate variables \cite{Heisenberg,Kennard},  such as position $\hat x$ and momentum  $\hat p$, which was later generalized to a rotation-invariant form by Schrödinger \cite{Schrodinger} and Robertson \cite{Robertson}. A different kind of uncertainty relations, originated by Bialynicki-Birula and Mycielski \cite{Birula} again for canonically-conjugate variables, relies on Shannon entropy instead of variances as a measure of uncertainty (it was later on developed for discrete observables of finite-dimensional systems \cite{Deutsch,Kraus,MaassenUffink}, but we restrict to continuous-variable observables here).

This entropic formulation of the uncertainty principle has recently attracted much attention in quantum information sciences  because entropies are the natural quantities of interest in this area (see \cite{Birula2,Coles2} for a survey). In particular, an extended version of the entropic uncertainty relation was derived, where some available quantum side-information (e.g., a quantum memory) is taken into account \cite{Renes,Berta}. It expresses the tradeoff between the information that two parties may have on non-commuting observables, which is of particular relevance to quantum key distribution. A variant version of this uncertainty relation formulated in terms of smooth entropies \cite{Tomamichel1} indeed provides a very useful tool for finite-key security analysis \cite{Tomamichel2}, going beyond asymptotic proofs. In the special case of continuous-variable quantum key distribution, the original entropic uncertainty relation \cite{Birula} was first applied to proving the optimality of Gaussian individual attacks at the asymptotic key limit \cite{GrosshansCerf}. More recently, a finite-key analysis for certain continuous-variable protocols was performed based on the smooth-entropy formalism extended to infinite dimensions \cite{Furrer}.

Entropic uncertainty relations find other applications, for example, in the context of  separability criteria. The Duan-Simon separability criteria \cite{duan,simon} based on variances for continuous-variables systems can be reformulated with entropies \cite{walborn},  yielding a more sensitive detection of entanglement in some cases. Even more generally, a deep conceptual link between the entropic uncertainty relation and the wave-particle duality has been pointed out \cite{Coles}, which emphasizes the pivotal role of entropies in the uncertainty principle.

In this article, we investigate whether tighter entropic uncertainty relations can be derived, which, by taking correlations into account, are saturated for \textit{all} Gaussian pure states  (in analogy with the Schrödinger-Robertson uncertainty relations). To reach this goal, we make use of the \textit{entropy power}, which is  a standard notion in Shannon information theory for real-valued signals.  In Section~II, we first review variance- and entropy-based uncertainty relations, and then define what we coin the \textit{entropy-power uncertainty relation} for a pair of canonically-conjugate variables, namely
$
N_x \, N_p \ge ( \hbar / 2 )^2,
$
where $N_x$ and $N_p$ are entropy powers.
It trivially implies the Heisenberg relation as a simple consequence of the definition of entropy power (actually, they coincide for Gaussian states). Then, in Section~III, we find an extended form of the entropy-power uncertainty relation, which is stronger than the regular form for rotated variables as it builds on the covariance matrix $\gamma$. It reads as 
\begin{equation}
N_x \, N_p  \ge  {\sigma_x^2 \, \sigma_p^2 \over  | \gamma |}  ~ ( \hbar / 2 )^2 
\label{conjecture-bis-intro}
\end{equation}
where $\sigma_x^2$ and $\sigma_p^2$ are variances.
It is partially proven by making use of variational calculus, supplemented with some natural assumption on the concavity of the uncertainty functional. 
We also find an extended version of the above entropy-power (or entropic) uncertainty relation that is valid for $n$ modes and is saturated for all $n$-mode Gaussian pure states (the proof is given in Appendix B). In Appendix A, we conduct extensive numerical tests in order to illustrate the validity of our extended uncertainty relations and conjectured concavity.

\section{From variance-based to entropy-power uncertainty relations}

\subsection{Variance-based uncertainty relations}

\begin{figure}[h!]
 \begin{tikzpicture}
 \draw[very thick, ->] (0, 0) -- (1.7, 0) node[right]{$x$};
 \draw[very thick, ->] (0, 0) -- (0, 1.7) node[left]{$p$};
 \draw[blue, very thick, ->] (0, 0) -- (1.53, 1) node[right]{$x_\theta$};
 \draw[blue, very thick, ->] (0, 0) -- (-1, 1.53) node[left]{$p_\theta$};
 \draw (1, 0) arc (0: 30: 1cm);
 \node at (1.1,0.25) {$\theta$}; 
 \end{tikzpicture}
 	\caption{\label{axesprinc} Principal axes ($x_\theta,p_\theta$) of the covariance matrix $\gamma$, defined in such a way that $\sigma_{x_\theta\, p_\theta}=0$.}
 \end{figure}
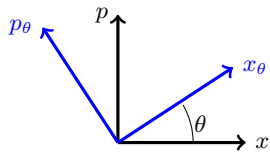

The original uncertainty relation, due to Heisenberg \cite{Heisenberg} and Kennard \cite{Kennard}, relies on the variances of $\hat x$ and $\hat p$. In the rest of this paper, we use quantum optics notations, so variables $\hat x$ and $\hat p$ stand for the quadrature components of a bosonic field (but they can, of course, also be viewed as the position and momentum variables of a mechanical degree of freedom). Using $[\hat x, \hat p]=i \hbar$, the Heisenberg uncertainty relation is written as
\begin{equation}
\sigma_x^2 \, \sigma_p^2 \ge ( \hbar / 2 )^2
\label{heisenberg}
\end{equation}
with variances $\sigma_x^2=\langle (\hat x-\bar x)^2 \rangle$ and $\sigma_p^2=\langle (\hat p-\bar p)^2 \rangle$, and mean values  $\bar x=\langle \hat x \rangle$ and $\bar p=\langle \hat p \rangle$. Here, $\langle \cdot \rangle \equiv\textrm{Tr}(\hat\rho \, \cdot)$ denotes the expectation value of ``$\cdot$" in quantum state $\hat \rho$. With this convention, the vacuum noise variances are $\sigma_{x,\textrm{vac}}^2 = \sigma_{p,\textrm{vac}}^2 = \hbar/2$.
Relation (\ref{heisenberg}) is invariant under  $(x,p)$-displacements in phase space, since it only depends on central moments (esp. second-order moments of the deviations from the means). Furthermore, it is saturated by all pure Gaussian states provided that they are squeezed in the $x$ or $p$ direction only.
More precisely, if we define the covariance matrix
\begin{equation}
\gamma = \begin{pmatrix}
\sigma_x^2 & \sigma_{xp} \\ 
\sigma_{xp} & \sigma_p^2 
\end{pmatrix} 
\end{equation}
where $\sigma_{xp} =  \langle \{ \hat x,\hat p \} \rangle /2 - \bar x \bar p$ is the symmetrized form of the central second-order cross moment,
we note that the Heisenberg relation is saturated for pure Gaussian states provided the principal axes of $\gamma$ are aligned with the $x$- and $p$-axes, namely  $\sigma_{xp}=0$.  The principal axes are the $x_\theta$- and $p_\theta$-axes for which $\sigma_{x_\theta\, p_\theta}=0$, where 
\begin{eqnarray}
\hat x_\theta = \cos \theta \, \hat x + \sin \theta \, \hat p  \qquad
\qquad \qquad \hat p_\theta = - \sin \theta \, \hat x + \cos \theta \, \hat p 
\end{eqnarray}
are obtained by rotating $x$ and $p$ by an angle $\theta$ as shown in Figure \ref{axesprinc}.
 
The Heisenberg relation was improved by Schrödinger and Robertson \cite{Schrodinger, Robertson} by taking into account the anticommutator between the observables\footnote{For any pair of observables $\hat A$ and $\hat B$, the generalized form of the uncertainty relation is
$\sigma_A^2 \, \sigma_B^2 \ge {1\over 4} |\langle[\hat A,\hat B]\rangle |^2 +  {1\over 4} | \langle \{\hat A',\hat B' \}\rangle |^2 $,
where $\hat A' = \hat A - \langle \hat A \rangle$ and $\hat B' = \hat B - \langle \hat B \rangle$.}. 
For two canonically-conjugate variables $\hat x$ and $\hat p$, it is written as
\begin{equation}
| \gamma | \ge ( \hbar / 2 )^2
\label{schrod-robert}
\end{equation}
where $|\gamma|=\sigma_x^2\sigma_p^2-\sigma_{xp}^2$ is the determinant of the covariance matrix. Importantly, relation (\ref{schrod-robert}) is saturated by all pure Gaussian states, regardless of the orientation of the principal axes of the covariance matrix. Thus, this uncertainty relation has the nice property of being invariant under all Gaussian unitary transformations (displacements and symplectic transformations).

\subsection{Entropy-based uncertainty relations}

The uncertainty principle may also be expressed using the entropy as a measure of uncertainty. In particular, Bialynicki-Birula and Mycielski \cite{Birula} proved the following \textit{entropic  uncertainty relation}
\begin{equation}
h(x)+h(p) \ge \ln(\pi e \hbar)
\label{birula}
\end{equation}
where $h(x)$ and $h(p)$ are the Shannon differential entropies of the $x$- and $p$-quadratures, namely
\begin{eqnarray}
h(x)=-\int W_x(x) \, \ln W_x(x) \, \mathrm dx,\qquad \qquad
h(p)=-\int W_p(p) \, \ln W_p(p) \, \mathrm dp.
\label{diffentropy}
\end{eqnarray}
Here, $W_x(x)=\int W(x,p) \, \mathrm dp$ and $W_p(p)=\int W(x,p) \, \mathrm dx$ denote the marginals of the Wigner function of state $\hat \rho$,
\begin{equation}
W(x,p)=\frac{1}{2 \pi \hbar} \int_{-\infty }^{\infty } e^{-\frac{i p y}{\hbar}}\, \mean{x+y/2 | \hat \rho | x-y/2}\, \mathrm dy
\end{equation}
so they are classical probability densities. 

Note that Eq. (\ref{birula}) may look wrong at first sight as we take the logarithm of a quantity with dimension $\hbar$. This may be viewed as a feature of the differential entropy itself, since we have a similar issue in Eq. (\ref{diffentropy}) itself, but the problem actually cancels out in Eq. (\ref{birula})  
since we have dimension $\hbar$ on both sides of the equality. More rigorously, Eq. (\ref{birula}) may be understood as the limit of a discretized version of the entropic uncertainty relation, with a discretization step tending to zero \cite{Birula2}. This problem was absent in the original expression of this uncertainty relation \cite{Birula} because the variable $k=p/\hbar$ was considered instead of $p$, giving $h(x)+h(k) \ge \ln(\pi e)$. Being aware of this slight abuse of notation, we prefer to keep $\hbar$ in the rest of this paper.

Just as the Heisenberg uncertainty relation, Eq. (\ref{birula}) is saturated by pure Gaussian states whose principal axes are aligned with the $x$- and $p$-axes (i.e.,  $\sigma_{xp}=0$). Indeed, for a Gaussian-distributed variable $x_G$ of variance $\sigma_x^2$ and a Gaussian-distributed variable $p_G$ of variance $\sigma_p^2$, we have
\begin{eqnarray}
h(x_G) = {1\over 2} \ln(2\pi e \sigma_x^2) \qquad \qquad
h(p_G) = {1\over 2} \ln(2\pi e \sigma_p^2).
\end{eqnarray}
Hence, summing up these two entropies and using the fact that  $\sigma_x^2 \, \sigma_p^2 = (\hbar/2)^2$ for any pure Gaussian state whose principal axes are aligned with the  $x$- and $p$-axes, we get
\begin{eqnarray}
h(x_G) + h(p_G) &=&\ln(\pi e \hbar).
\end{eqnarray}

Remark that we may also re-express the entropic uncertainty relation in terms of relative entropies\footnote{The relative entropy between two probability densities $f(x)$ and $g(x)$ is defined as $D(f || g) = \int  f(x) \, \ln(f(x)/g(x))  \, \mathrm dx$. It exhibits the property that $D(f || g)\ge 0$, and $D(f || g)= 0$ if and only if $f(x)=g(x)$, $\forall x$ (almost everywhere).}. More precisely, using a measure of non-Gaussianity that relies on the relative entropy \cite{Genoni}, we have 
\begin{eqnarray}
D(x||x_G) = h(x_G) - h(x)  \ge 0 \qquad\qquad
D(p||p_G) = h(p_G) - h(p) \ge 0
\end{eqnarray}
so that  the entropic uncertainty relation is equivalent to
\begin{equation}
D(x||x_G) + D(p||p_G) \le  \ln \left( {\sigma_x \sigma_p \over \hbar / 2  } \right).
\label{EURwithD}
\end{equation}
We see immediately that if the Heisenberg relation is saturated, $\sigma_x \sigma_p = \hbar / 2$, then 
$D(x||x_G)=D(p||p_G)=0$, which means that the $x$- and $p$-quadratures must both be Gaussian distributed. Thus, as emphasized in ref.~\cite{Son}, the entropic uncertainty relation may also been viewed as an improved version of the Heisenberg relation where the lower bound is lifted up by exploiting an entropic measure of the non-Gaussianity of the state, namely
\begin{equation}
\sigma_x^2 \, \sigma_p^2 \ge  ( \hbar / 2 )^2 \,  e^{2 D(x||x_G) + 2 D(p||p_G)} .
\label{EURwithDbis}
\end{equation}

\subsection{Entropy-power uncertainty relations}

We will show now that it is possible to rewrite the entropic uncertainty relation in a form similar to the one expressed in terms of variances, provided we make use of the notion of entropy power\footnote{The entropy power $N(X)$ of a real-valued random variable $X$ is defined as the variance of a Gaussian-distributed random variable  having the same entropy as $X$ (the mean of $X$ plays no role since the entropy is translation-invariant). Since the distribution with highest entropy for a given variance is the Gaussian distribution, $N(X)\le \sigma_X^2$,  the equality being reached if and only if $X$ is Gaussian distributed.}. The entropy power of the $x$- and $p$-quadratures are defined as
\begin{eqnarray}
N_x = {1\over 2\pi e} \, e^{2\, h(x)} \qquad\qquad
N_p = {1\over 2\pi e} \, e^{2\, h(p)} \,  ,
\end{eqnarray}
and we have  $N_x=\sigma_x^2$ and $N_p=\sigma_p^2$ if and only if the $x$- and $p$-quadratures are Gaussian distributed.
Thus, Eq. (\ref{birula}) can be simply reexpressed as
\begin{equation}
N_x \, N_p \ge ( \hbar / 2 )^2  \,  ,
\label{EPUR}
\end{equation}
which is what we call an \textit{entropy-power uncertainty relation} for a pair of canonically-conjugate variables, as presented in the introduction. It closely resembles the Heisenberg relation (\ref{heisenberg}), but with entropy powers instead of variances.

Since $N_x \le \sigma_x^2$ and $N_p \le \sigma_p^2$, which reflects the fact that the Gaussian distribution maximizes the entropy for a fixed variance, we have the chain of inequalities
\begin{equation}
\sigma_x^2 \, \sigma_p^2 \ge N_x \, N_p \ge ( \hbar / 2 )^2.
\end{equation}
Hence, the entropy-power uncertainty relation implies the Heisenberg uncertainty relation, and they coincide for Gaussian $x$- and $p$-distributions (this was already mentioned in \cite{Birula}). This can also be connected to relative entropies as a measure of non-Gaussianity. From the definition of $N_x$ and $N_p$, we get
\begin{eqnarray}
h(x) = {1\over 2} \ln(2\pi e N_x) \qquad\qquad
h(p) = {1\over 2} \ln(2\pi e N_p)  \,  
\end{eqnarray}
which implies that 
\begin{eqnarray}
D(x||x_G) =  {1\over 2} \ln \left( {\sigma_x^2 \over N_x } \right)  \qquad\qquad
D(p||p_G) =  {1\over 2} \ln \left( {\sigma_p^2 \over N_p } \right) 
\end{eqnarray}
or equivalently
\begin{eqnarray}
 \sigma_x^2  = N_x  \, e^{2\, D(x||x_G)}  \qquad\qquad   \sigma_p^2  = N_p  \, e^{2\, D(p||p_G)}    \,  .
\end{eqnarray}
It is clear that Eq.~(\ref{EPUR}) becomes more stringent than Eq.~(\ref{heisenberg}) as soon as we deviate from a Gaussian state.

\section{Extended forms of entropic uncertainty relations}

\subsection{Motivation}

Our goal is to address the problem that, unlike the Schrödinger-Robertson uncertainty relation, the entropic uncertainty relation (\ref{birula}) --
or equivalently the entropy-power uncertainty relation (\ref{EPUR}) -- is not saturated by all pure Gaussian states but only by those whose principal axes are aligned with the $x$- and $p$-axes. In other words, we would like to make Eq. (\ref{birula}) or (\ref{EPUR}) depend on the possible correlations between $x$ and $p$ (as witnessed, for instance, by $\sigma_{xp}\ne 0$). Ideally, the new inequality should have the property of being invariant under all Gaussian unitary transformations (displacements and symplectic transformations) and being saturated by all pure Gaussian states, regardless of the orientation of the principal axes.

A first natural idea is to make use of the joint differential entropy, which is defined as
\begin{equation}
h(x,p)=-\int f(x,p)\ln f(x,p) \, \mathrm dx \, \mathrm dp
\end{equation}
where $f(x,p)$ is the joint probability density of the random variables $x$ and $p$. The joint entropy can also be expressed as $h(x,p)=h(x)+h(p)-I(x{\rm :}p)$ where $I(x{\rm :}p)\ge 0$ is the mutual information.  Thus, one may think of improving the entropic uncertainty relation (\ref{birula}) by replacing $h(x)+h(p)$ with $h(x,p)$. Moving the mutual information $I(x{\rm :}p)$ on the right-hand side of the inequality, it thus corresponds to an improvement of the lower bound.
Moreover, $h(x,p)$ has the invariance property that we seek. Indeed, if we transform the coordinates according to $(x'\,\, p')^T=S \cdot (x \,\, p)^T$, where $S$ is the transformation matrix, the joint differential entropy transforms as \cite{Cover}
\begin{equation}
h(x',p')=h(x,p)+\ln |S|.
\end{equation}
Thus, if $S$ corresponds to a symplectic transformation, $|S|=1$, then the joint differential entropy remains invariant. Of course,  $h(x,p)$ is also invariant under ($x$,$p$)-displacement, so it looks like a good uncertainty functional.

However, we deal with quantum states, so the Wigner function $W(x,p)$ is not a genuine probability density and may admit negative values. Hence, the joint differential entropy of $W(x,p)$ is not always defined (one would need to compute the logarithm of negative values), and so is the mutual information $I(x{\rm :}p)$. Nevertheless, we conjecture that the joint differential entropy obeys  a valid uncertainty relation if we restrict to states admitting a Wigner function that is non-negative everywhere, namely
\begin{equation}
h(x,p)\geq \ln(\pi e \hbar)\,\,\,\,\,\,\,\,  \forall {\rm ~states~s.t.~} W(x,p)\geq0.
    \label{conj1form1}
\end{equation}
This conjecture can equivalently be written as
  \begin{equation}
  h(x)+h(p)\geq \ln(\pi e \hbar)+I(x{\rm :}p)     \,\,\,\,\,\,\,\,  \forall {\rm ~states~s.t.~} W(x,p)\geq0,
  \label{conj1form2}
  \end{equation}
which is an improvement over Eq. (\ref{birula}) since $I(x{\rm :}p)\ge 0$.

A difficulty, however, is related to the fact that characterizing the set of states with positive Wigner functions is not an easy task \cite{Brocker}. In addition, for states admitting negative Wigner functions, Eq.~(\ref{conj1form1}) or (\ref{conj1form2}) is useless. In Appendix A, we run numerics to check the validity of Eq.~(\ref{conj1form2}), and provide examples of states where Eq.~(\ref{conj1form2}) gives a slightly better bound than Eq.~(\ref{birula}) although the correlation between $x$ and $p$ is not accessible via the second-order moments ($\sigma_{xp}=0$) but via the mutual information $I(x{\rm :}p)$ only.
  
 \subsection{Tight entropy-power uncertainty relation involving the covariance matrix}

Equations (\ref{conj1form1}) or (\ref{conj1form2}) are not valid for states with negative Wigner functions, but they give us a hint on how to proceed in order to derive an entropic uncertainty relation that is valid for all states and takes correlations into account. While the  joint entropy and mutual information are not defined for all states, they are well defined for Gaussian states (since their Wigner function is always positive). In particular, the Gaussian mutual information is expressed as a function of the covariance matrix,
\begin{equation}
	I_G(x{\rm :}p)=\frac{1}{2}\ln\left(\sigma_x^2\sigma_p^2 / |\gamma| \right) \ge 0.
\end{equation}
We obtain our tight entropic uncertainty relation simply by substituting $I(x{\rm :}p)$ with $I_G(x{\rm :}p)$ in Eq. (\ref{conj1form1}), namely
\begin{equation}
h(x)+h(p) -  {1\over 2} \ln \left(  \sigma_x^2 \sigma_p^2  / |\gamma|  \right)  \ge \ln(\pi e \hbar). 
\label{conjecture}
\end{equation}
We will show below (under some assumptions) that this inequality holds for all states, regardless of whether the Wigner function is positive everywhere or not. Unlike Eq. (\ref{conj1form1}),  however, it is not invariant under rotations. Note that $I_G(x{\rm :}p)$ vanishes if the principal axes of the covariance matrix are the $x$- and $p$-axes, i.e. $\sigma_{xp}=0$, so that Eq.~(\ref{conjecture}) reduces to the regular entropic uncertainty relation (\ref{birula}) in this case.

As before, it is useful to rewrite our new relation in terms of entropy powers (as we presented it in the introduction), resulting in
\begin{equation}
N_x \, N_p  \ge  {\sigma_x^2 \, \sigma_p^2 \over  | \gamma |}  ~ ( \hbar / 2 )^2 
\label{conjecture-bis}
\end{equation}
which can be viewed as an improved version of the entropy-power uncertainty relation (\ref{EPUR}), where the lower bound $(\hbar/2)^2$ is lifted up when the principal axes differ from the $x$- and $p$-axes ($\sigma_{xp}\ne 0$). If the principal axes correspond to the $x$- and $p$-axes, we recover Eq. (\ref{EPUR}).
Alternatively, we may also reexpress our new relation as
\begin{equation}
 {N_x \, N_p\over \sigma_x^2 \, \sigma_p^2} ~  | \gamma | \ge  ( \hbar / 2 )^2.
\end{equation}
Then, using $N_x \le \sigma_x^2$ and $N_p \le \sigma_p^2$, we see that our tight entropy-power inequality (\ref{conjecture-bis}) implies  the Schr\"odinger-Robertson uncertainty relation, namely
\begin{equation}
 | \gamma | \ge  {N_x \, N_p\over \sigma_x^2 \, \sigma_p^2} ~  | \gamma | \ge  ( \hbar / 2 )^2.
\end{equation}
These two inequalities coincide for Gaussian $x$- and $p$-distributions. Furthermore, they are both saturated for pure Gaussian states regardless the orientation of the principal axes (since $ | \gamma | = ( \hbar / 2 )^2$ and $N_x=\sigma_x^2$, $N_p=\sigma_p^2$).

In addition, we may reexpress Eq. (\ref{conjecture-bis}) as
\begin{equation}
 | \gamma | \ge  {\sigma_x^2 \, \sigma_p^2 \over N_x \, N_p}  ~ ( \hbar / 2 )^2 
\end{equation}
which can be viewed as an improved version of the Schr\"odinger-Robertson uncertainty relation where  the lower bound $(\hbar/2)^2$ is lifted up when the $x$- and $p$-distributions deviate from Gaussian distributions.  In terms of non-Gaussianity measures based on relative entropies, it transforms into
\begin{equation}
D(x||x_G) + D(p||p_G) \le  \ln \left( { \sqrt{|\gamma|} \over \hbar / 2 } \right).
\label{conjecture2}
\end{equation}
which is the counterpart of Eq. (\ref{EURwithD}) but having replaced $\sigma_x^2 \, \sigma_p^2$ with $ |\gamma|$, just as we do when going from the Heisenberg to the Schr\"odinger-Robertson relation. It also corresponds to a stronger version of Eq. (\ref{EURwithDbis}), which reads
\begin{equation}
|\gamma|^{1/2} \ge  ( \hbar / 2 ) \,  e^{D(x||x_G) + D(p||p_G)} .
\end{equation}

To be complete, let us mention that we can express our tight entropic uncertainty relation (\ref{conjecture})  as
\begin{equation}
h(x)+h(p)\geq h(x_G)+h(p_G)+\ln(\mu_G)
\end{equation}
where $x_G$ ($p_G$) is Gaussian distributed with variance $\sigma_x^2$ ($\sigma_p^2$) and $\mu_G = {\rm tr} \rho_G^{\,\,\,2}$ is the purity of the Gaussian state $\rho_G$ associated to the covariance matrix $\gamma$.

Finally, note that our conjectured rotation-invariant uncertainty relation (\ref{conj1form1}) based on the joint entropy is obviously equivalent to Eq. (\ref{conjecture}) for Gaussian states, so in both cases the bound is reached by any pure Gaussian state, regardless of the orientation of the principal axes. Thus, by taking the exponential of the joint entropy $h(x,p)$ and using the fact that the maximum entropy is reached for a Gaussian distribution, a similar derivation shows that relation (\ref{conj1form1}) also implies the Schrödinger-Robertson uncertainty relation.

\subsection{Partial proof of relation (\ref{conjecture})}

We now give a partial proof of our tight entropic uncertainty relation (\ref{conjecture}) by use of a variational method, in analogy to the procedure used in ref. \cite{Hall} to prove a noise-dependent entropic uncertainty relation. More precisely, we will prove that any squeezed vacuum state rotated by an arbitrary angle is a local minimum of the uncertainty functional 
\begin{equation}
F(\hat\rho) = h(x)+h(p) -  {1\over 2} \ln \left(   \sigma_x^2 \sigma_p^2  / |\gamma|  \right).
\label{uncertainty-functional}
\end{equation}
Since $F(\hat\rho)$ is invariant under $(x,p)$-displacements, it will imply that all Gaussian pure states are similarly local minima. We assume that these are the unique solutions of our minimization problem. By assuming that the uncertainty functional $F(\hat\rho)$ is concave in $\rho$, which we have verified numerically in Section IV, we also conclude that relation (\ref{conjecture}) is valid for mixed states as well. We know that this situation prevails for the regular entropic uncertainty relation (\ref{birula}) as well as for our conjectured relation (\ref{conj1form1}), so the above assumptions (unicity and concavity) are very natural.

Let us seek for a pure state $\ket{\psi}$ that minimizes the functional $F(\proj{\psi})$. For this, we use the Lagrange multiplier method and insert the normalization of  $\ket{\psi}$ as a constraint. Since $F(\proj{\psi})$ is invariant under displacements, we may also impose with no loss of generality the constraint that mean values vanish, $\langle \hat x \rangle = \langle \hat p \rangle = 0$. We define
\begin{equation}
J=F(\proj{\psi})+\lambda( \mean{\psi|\psi}-1) +\mu \mean{\psi|\hat{x}|\psi}+\nu \mean{\psi|\hat{p}|\psi}
\end{equation}
where $\lambda$, $\mu$ and $\nu$ are Lagrange multipliers. Since we impose the state to be normalized and centered on zero, we can express the second-order moments as $\sigma_x^2=\mean{\psi|\hat{x}^2|\psi}$, $\sigma_p^2=\mean{\psi|\hat{p}^2|\psi}$,  and  $\sigma_{xp}=\frac{1}{2}\mean{\psi|\{\hat{x},\hat{p}\}
|\psi}$, so that we may replace the  functional $F(\proj{\psi})$ in $J$ by 
\begin{eqnarray}
   \tilde F(\proj{\psi}) =h(x)+h(p)  -\frac{1}{2}\ln\left(\frac{\mean{\psi|\hat{x}^2|\psi}\mean{\psi|\hat{p}^2|\psi}}{\mean{\psi|\hat{x}^2|\psi}\mean{\psi|\hat{p}^2|\psi}-\frac{1}{4}\mean{\psi|\{\hat{x},\hat{p}\}|\psi}^2}\right).
\end{eqnarray}
Now, in order to solve the variational equation
\begin{equation}
\frac{\partial J}{\partial \bra{\psi}}=0
\label{variational}
\end{equation}
we start by expressing the variational derivative of each term of $J$ separately. The first term gives
\begin{eqnarray}
\frac{\partial h(x)}{\partial \bra{\psi}}&=&\frac{\partial}{\partial \bra{\psi}}\left(-\int W_x(x)\ln W_x(x) \mathrm dx \right)\nonumber\\
&=&\frac{\partial}{\partial \bra{\psi}}\left(-\int \mean{\psi|x}\mean{x|\psi}\ln( \mean{\psi|x}\mean{x|\psi}) \mathrm dx \right)\nonumber\\
&=&-\left( \ln W_x(\hat{x}) +1\right) \ket{\psi}
\end{eqnarray}
and, similarly,  the second term gives
\begin{equation}
\frac{\partial h(p)}{\partial \bra{\psi}}=-\left(  \ln W_p(\hat{p}) +1\right) \ket{\psi}.
\end{equation}
For the third term, we use
\begin{eqnarray}
    \frac{\partial }{\partial \bra{\psi}} \ln\left(
\frac{\mean{\psi|\hat{x}^2|\psi}\mean{\psi|\hat{p}^2|\psi}}
{\mean{\psi|\hat{x}^2|\psi}\mean{\psi|\hat{p}^2|\psi}-\frac{1}{4}\mean{\psi|\{\hat{x},\hat{p}\}|\psi}^2}
\right)    =\left[  \frac{\hat{x}^2}{\sigma_x^2} +\frac{\hat{p}^2}{\sigma_p^2}
 -\frac{\hat{x}^2\sigma_p^2+\hat{p}^2\sigma_x^2- \{\hat{x},\hat{p}\}\sigma_{xp}}{|\gamma|}
\right]  \ket{\psi}
\end{eqnarray}
while the last terms give
\begin{eqnarray}
\frac{\partial}{\partial \bra{\psi}} \bigg(  \lambda( \mean{\psi|\psi}-1) + \mu \mean{\psi|\hat{x}|\psi} + \nu \mean{\psi|\hat{p}|\psi}  \bigg)  
= \left(\lambda+\mu\hat{x}+\nu\hat{p}\right)\ket{\psi}.
\end{eqnarray}
Putting all this together, the variational equation (\ref{variational}) can be rewritten as an eigenvalue equation for $ \ket{\psi}$,
\begin{eqnarray}
 \bigg[-\ln W_x(\hat{x}) -\ln W_p(\hat{p}) -2+\lambda+\mu\hat{x}+\nu\hat{p}   - \frac{\hat{x}^2}{2 \sigma_x^2} -\frac{\hat{p}^2}{2 \sigma_p^2}
 +\frac{\hat{x}^2\sigma_p^2+\hat{p}^2\sigma_x^2- \{\hat{x},\hat{p}\}\sigma_{xp}}{2 |\gamma|}
  \bigg]  \ket{\psi}=0 . \hspace{0.8cm}  
\label{eigenequ}
\end{eqnarray}
Let us check that Eq. (\ref{eigenequ}) is verified by $\ket{\psi}=\hat S\ket{0}$, that is, by a squeezed vacuum state with $\hat S~=~\exp\{\frac{1}{2}(z^*\hat{a}^2-z\hat{a}^{\dag^2})\}$, where $z=r e^{i \phi}$ is a complex number. For such a state, the marginals of the Wigner functions are given by
\begin{equation}
W_x(x)=(2\pi\sigma_x^2)^{-{1\over 2}} \, e^{-\frac{x^2}{2\sigma_x^2}},  \qquad\qquad  W_p(p)= (2\pi\sigma_p^2)^{-{1\over 2}} \,  e^{-\frac{p^2}{2\sigma_p^2}},
\end{equation}
so that
\begin{equation}
\ln W_x(\hat{x}) + \ln W_p(\hat{p}) =-\ln(2\pi \sigma_x \sigma_p)-\frac{\hat{x}^2}{2\sigma_x^2}-\frac{\hat{p}^2}{2\sigma_p^2}.
\end{equation}
Hence, we can simplify the eigenvalue equation as 
\begin{equation}
\left[\ln(2\pi \sigma_x \sigma_p)-2+\lambda+\mu\hat{x}+\nu\hat{p}+\hat{A}\right]\ket{\psi}=0
\end{equation}
where we have defined the operator
\begin{eqnarray}
\hat{A}= \frac{\hat{x}^2\sigma_p^2+\hat{p}^2\sigma_x^2- \{\hat{x},\hat{p}\}\sigma_{xp}}{2 |\gamma|} = {1\over 2} \begin{pmatrix} \hat{x} & \hat{p} \end{pmatrix} \, \gamma^{-1} \, \begin{pmatrix} \hat{x} \\ \hat{p} \end{pmatrix} .
\end{eqnarray}
Let us now compute the action of $\hat{A}$ on the squeezed vacuum state, that is, $\hat{A}\ket{\psi}=\hat{A} \hat{S} \ket{0}= \hat{S}(\hat{S}^{\dagger}\hat{A} \hat{S}) \ket{0}$. For this, we use the canonical transformation of $\hat x$ and $\hat p$ in the Heisenberg picture, namely
\begin{equation}
\begin{pmatrix}
{\hat S}^{\dagger} \hat{x}{\hat S} \\ {\hat S}^{\dagger} \hat{p} {\hat S}  \end{pmatrix}=M \begin{pmatrix}
\hat{x}\\\hat{p}
\end{pmatrix}
\label{canonical1}
\end{equation}
with
\begin{eqnarray}
M = \begin{pmatrix}  \cos\theta & - \sin \theta \\ \sin\theta & \cos\theta \end{pmatrix} 
\begin{pmatrix}  e^{-r} & 0\\ 0 & e^r \end{pmatrix} 
\begin{pmatrix}  \cos\theta & \sin \theta \\ -\sin\theta & \cos\theta \end{pmatrix}   = \begin{pmatrix}
\cosh r-\cos\phi\,\sinh r&-\sin\phi\,\sinh r\\-\sin\phi\,\sinh r&\cosh r+\cos\phi\,\sinh r
\end{pmatrix}   \hspace{0.8cm}
\end{eqnarray}
with $\phi=2\theta$.
The covariance matrix $\gamma$ of state $\ket{\psi}$ can be expressed with transformation $M$ applied onto the covariance matrix of the vacuum state $\gamma_\mathrm{vac}$, namely 
\begin{equation}
\gamma = M \gamma_\mathrm{vac} M^T  .
\label{canonical2}
\end{equation}
Using Eqs. (\ref{canonical1}) and (\ref{canonical2}), we get
\begin{eqnarray}
\hat{A}\ket{\psi}=\frac{1}{2}{\hat S} \begin{pmatrix}
\hat{x}& \hat{p} \end{pmatrix}M^T\gamma^{-1}M\begin{pmatrix}
\hat{x}\\\hat{p} \end{pmatrix}\ket{0}
=\frac{1}{2}{\hat S}   \begin{pmatrix}
\hat{x}& \hat{p} \end{pmatrix} \gamma_\mathrm{vac}^{-1} \begin{pmatrix}
\hat{x}\\\hat{p} \end{pmatrix}   \ket{0}
={\hat S}\ket{0} = \ket{\psi}
\end{eqnarray}
implying that the squeezed vacuum state $\ket{\psi}$ is an eigenvector of $\hat{A}$ with eigenvalue 1. Therefore, the eigenvalue equation can be written as 
\begin{eqnarray}
\left[\ln(2\pi \sigma_x \sigma_p)-1+\lambda+\mu\hat{x}+\nu\hat{p}\right]\ket{\psi}=0.
\end{eqnarray}
We can determine the value of $\lambda$ by multiplying this equation on the left by $\bra{\psi}$ and using the constraints  $ \mean{\psi|\psi}=1$ and $\mean{\psi|\hat{x}|\psi}=\mean{\psi|\hat{p}|\psi}=0$, namely
\begin{eqnarray}
\bra{\psi}   \left[\ln(2\pi \sigma_x \sigma_p)-1+\lambda+\mu\hat{x}+\nu\hat{p}\right]  \ket{\psi}
= \ln(2\pi \sigma_x \sigma_p)-1+\lambda  =0.
\end{eqnarray}
Therefore, state $\ket{\psi}$ is indeed a solution of our extremization problem if we set $\lambda=1-\ln(2\pi \sigma_x \sigma_p)$. We are left with equation
\begin{equation}
\left[\mu\hat{x}+\nu\hat{p}\right]\ket{\psi}=0
\end{equation}
which is satisfied if we set $\mu=\nu=0$. Summing up, we have proven that, with the appropriate choice of  $\lambda$, $\mu$ and $\nu$, the squeezed vacuum states (with arbitrary squeezing and rotation) are solutions of Eq. (\ref{eigenequ}), so they minimize our uncertainty functional $F(\proj{\psi})$. Since $F(\proj{\psi})$ is invariant under displacements, the displaced squeezed states are also solutions, so this result includes all pure Gaussian states. We  find the minimum value $\ln(\pi e \hbar)$ simply by evaluating $F$ for any of these states.

As mentioned above, this proof does not imply that the pure Gaussian states are the only minimum-uncertainty states, and we also need to assume the concavity of our uncertainty functional in order to extend the proof to mixed states. However, these are very natural assumptions, which are verified in the special case of states with $\sigma_{xp}=0$ since then we are back to the regular entropic uncertainty relation. Moreover, in Appendix A, we give strong numerical evidence that our tight entropic uncertainty relation is valid.
Other numerical tests also corroborate the concavity property of the uncertainty functional, while this property is proven in the special case when two states with the same covariance matrix are mixed.

\subsection{Generalization to $n$ modes}

In ref. \cite{Birula}, Bialynicki-Birula and Mycielski also extended the entropic uncertainty relation to $n$ modes, namely
\begin{equation}
h(\vec x)+h(\vec p)  \ge n\,\ln(\pi e \hbar)
\label{birulanmodes}
\end{equation}
where the joint differential entropies $h(\vec x)$ and $h(\vec p)$ are computed from the marginals of the Wigner functions $W_x(\vec x)$ and $W_p(\vec p)$, with $\vec x=(x_1,x_2,\cdots ,x_n)$ and $\vec p=(p_1,p_2,\cdots ,p_n)$.  In addition,  another $n$-mode uncertainty relation was expressed in ref. \cite{huang} for two observables $\hat A$ and $\hat B$ defined as linear combinations of  the $\hat x_i$ and $\hat p_i$ variables.

Naturally, both our entropic uncertainty relations can also be extended to $n$ modes. First, our conjectured rotation-invariant uncertainty relation based on the joint entropy  (\ref{conj1form1}) becomes
 \begin{equation}
h(\vec r)\geq n\,\ln(\pi e \hbar)\qquad\forall\,\text{states s.t. }W(\vec r)\geq0 
\label{joint-nmodes}
\end{equation}
where  $\vec r=(x_1,p_1,x_2,p_2,...,x_n,p_n)$. Here, the joint differential entropy $h(\vec r)$ is invariant under Gaussian $n$-mode unitaries (all symplectic transformations and displacements) and our conjectured uncertainty relation (\ref{joint-nmodes}) is saturated for all $n$-mode Gaussian pure states.
 
Second, our tight entropic uncertainty relation (\ref{conjecture}) can also be extended to 
\begin{equation}
h(\vec x)+h(\vec p)   -  {1\over 2} \ln \left(\frac{|\gamma_x||\gamma_p|}{ |\gamma| } \right) \ge n\,\ln(\pi e \hbar)
\label{nmodes}
\end{equation}
where the covariance matrix $\gamma$ is defined as
$\gamma_{ij}=\mathrm{Tr}[\hat \rho\, \{r_i,r_j\}]/2-\mathrm{Tr}[\hat\rho \, r_i]\mathrm{Tr}[\hat\rho \, r_j]$ and $\gamma_x$ ($\gamma_p$) is the reduced covariance matrix of the $x$ ($p$) quadratures.
The proof of this relation can be found in Appendix B (it is obtained following the same variational method as in the one-mode case). 
Equation (\ref{nmodes}) is again saturated by all $n$-mode Gaussian pure states, as we can easily check by using the fact that
\begin{equation}
h(\vec x)= {1\over2}\ln((2\pi e)^{n}|\gamma_x|)\qquad\qquad h(\vec p)= {1\over2}\ln((2\pi e)^{n}|\gamma_p|)
\label{entropyGauss}
\end{equation}
for Gaussian distributions,  while $|\gamma|=(\hbar/2)^{2n}$ for Gaussian pure states.

In particular, relation (\ref{nmodes})  is thus saturated by the two-mode vacuum squeezed state with covariance matrix 
\begin{equation}
\gamma={\hbar\over2}\begin{pmatrix}
\cosh 2r &0&\sinh 2r& 0 \\
0&\cosh 2r& 0 &-\sinh 2r\\
\sinh 2r&0&\cosh 2r&0\\
0&-\sinh 2r&0&\cosh 2r
\end{pmatrix} \, ,
\end{equation}
obtained by injecting an $x$-squeezed state and a $p$-squeezed state (both with a squeezing parameter $r$) on a balanced beam splitter. This is easy to check by computing the entropies with Eq.~(\ref{entropyGauss}) and using $|\gamma_x|=|\gamma_p|=(\hbar/2)^2$ and $|\gamma|=(\hbar/2)^4$. However, the regular entropic uncertainty relation (\ref{birulanmodes}) is already saturated for this state, which is expected since the state exhibits no $x$-$p$ correlations. More interestingly, the state resulting from two rotated squeezed states (one being rotated by $\pi/4$, the other by  $-\pi/4$) injected on a balanced beam splitter still saturates relation (\ref{nmodes}), while it does not any more saturate relation (\ref{birulanmodes}). Indeed, the covariance matrix of this state reads
\begin{equation}
\gamma={\hbar\over2}\begin{pmatrix}
\cosh 2r &0&0&-\sinh 2r \\
0&\cosh 2r&-\sinh 2r &0\\
0&-\sinh 2r&\cosh 2r&0\\
-\sinh 2r&0&0&\cosh 2r
\end{pmatrix}.
\end{equation}
so that we get $h(\vec x)+h(\vec p)=2\ln\left(\pi e\hbar\cosh 2r\right)  >  2\ln\left(\pi e\hbar\right)$.
But since $|\gamma_x|=|\gamma_p|= (\hbar/2)^2 \cosh^2 2r$ and $|\gamma|=(\hbar/2)^4$, we get
$-  {1\over 2} \ln \left(\frac{|\gamma_x||\gamma_p|}{ |\gamma| } \right)=-  2\ln \left(\cosh 2r \right)$, implying that relation (\ref{nmodes}) is saturated by this state.

In this context, it is also interesting to rewrite the tight entropic uncertainty relation (\ref{nmodes}) in term of entropy powers, defined this time for the joint entropy in $n$ dimensions, namely
\begin{equation}
N_x^{(n)}=\frac{1}{2\pi e}e^{{2\over n}h(\vec x)}\qquad\qquad 
N_p^{(n)}=\frac{1}{2\pi e}e^{{2\over n}h(\vec p)}
\end{equation}
Equation (\ref{birulanmodes}) then transforms into a $n$-mode entropy-power uncertainty relation 
\begin{equation}
N_x^{(n)} N_p^{(n)} \geq (\hbar/2)^2 \, ,
\end{equation}
which has the same form as relation (\ref{EPUR}) but for $n$ modes, while equation (\ref{nmodes}) transforms into a tight version of the $n$-mode entropy-power uncertainty relation 
\begin{equation}
N_x^{(n)} N_p^{(n)} \geq\left(\frac{|\gamma_x| \, |\gamma_p|}{|\gamma|}\right)^{1/n} \, (\hbar/2)^2 \, ,
\label{n-mode-conjecture-bis}
\end{equation}
which is the $n$-mode counterpart of Eq. (\ref{conjecture-bis}).

Here too, we can use the fact that the maximum entropy for a fixed covariance matrix is given by the Gaussian distribution, which implies that $N_x^{(n)} \leq |\gamma_x|^{1/n}$ and $N_p^{(n)} \leq|\gamma_p|^{1/n}$. Rewriting Eq. (\ref{n-mode-conjecture-bis}) as 
\begin{equation}
\frac{\left( N_x^{(n)} N_p^{(n)} \right)^n}{|\gamma_x| \, |\gamma_p|} \, |\gamma| \geq (\hbar/2)^{2n}   \, ,
\end{equation}
we then see that the  $n$-mode entropy-power uncertainty relation implies the standard (variance-based) $n$-mode  uncertainty relation, namely
\begin{equation}
|\gamma|\geq\frac{\left( N_x^{(n)} N_p^{(n)} \right)^n}{|\gamma_x|\, |\gamma_p|} \, |\gamma| \geq  (\hbar/2)^{2n} .
\end{equation}

\section{Conclusion}

We have shown that the entropic uncertainty relation derived by Bialynicki-Birula and Mycielski can can be expressed as an entropy-power uncertainty relation, 
which makes a straightforward connection with Heisenberg uncertainty relation 
: the variances in the latter are simply replaced with entropy powers in the former. Moreover,
the entropic version of the uncertainty relation
implies the variance-based one 
as a consequence of the fact that the entropy power of a variable cannot exceed its variance. Then, we have found a tighter form of the entropic uncertainty relation, 
 which takes the correlation between the $x$- and $p$-variables into account. It can also be expressed as a tighter entropy-power uncertainty relation, Eq.~(\ref{conjecture-bis-intro}), and is saturated for all pure Gaussian states. It is the entropic counterpart of the Schrödinger-Robertson uncertainty relation, 
 which it implies.
We have provided a partial proof of Eq.~(\ref{conjecture-bis-intro}) based on variational calculus together with some reasonable assumptions, and have provided, in the Appendix A, strong numerical evidence that it is correct. Interestingly, this tighter entropic and entropy-power uncertainty relations can be extended  to $n$ modes,
and all the above-mentioned properties remain true. 
Our main result was inspired from another conjectured uncertainty relation involving the joint entropy, Eq.~(\ref{conj1form1}), which is more elegant (it is explicitly invariant under all Gaussian unitaries -- displacements, squeezing, and rotations) but is only defined for states with a non-negative Wigner function. 
We have numerically verified its validity, but leave its proof for further work. Its $n$-mode extension is also straightforward.

Possible applications of these new entropic uncertainty relations include the elaboration of stronger separability criteria for continuous-variable systems. Both variance- and entropy-based uncertainty relations can be translated into a sufficient entanglement condition (a necessary and sufficient condition for Gaussian states) as they can be used to express a condition on the physicality of the partially-transposed state \cite{duan, simon, walborn}. For example, in ref. \cite{hertz} it was shown that an uncertainty relation that is tight for all Fock states \cite{katerina} yields an entanglement criterion that enables the detection of certain non-Gaussian entangled states whose entanglement remains undetected by the Duan-Simon criterion. Thus, a natural direction for further work would be to exploit our tighter entropic uncertainty relations in order to improve our tools for discriminating entangled from separable states in continuous-variable quantum systems.

\medskip 
\noindent {\it Note}: The current paper was presented at the 23rd Central European Conference on Quantum Optics (CEWQO 2016), Kolymbari, Greece, June  2016. After completion of this work, we learned about an independent work where the entropy power is mentioned in the context of uncertainty relations \cite{Jizba}.

\medskip
\noindent {\it Acknowlegments}: We thank Emmanouïl Grigoriou for performing numerical simulations during a research internship at QuIC, ULB, in Summer 2016.
This work was supported by the F.R.S.-FNRS Foundation under Project No. T.0199.13 and by the Belgian Federal IAP program under Project No. P7/35 Photonics@be. A.H. acknowledges financial support from the F.R.S.-FNRS Foundation and M.G.J. acknowledges financial support from the FRIA foundation.


\section*[appendix]{APPENDIX A : Numerical tests}
\def\thesubsection{\arabic{subsection}}

\subsection{Numerical tests of the uncertainty relation  (\ref{conj1form1})}

We have not been able to find an analytical proof of our conjectured rotation-invariant uncertainty relation (\ref{conj1form1}) based on the joint entropy, so we have turned to numerical tests. Since relation (\ref{conj1form1}) is restricted to states with positive Wigner functions, we have tested, in particular, passive states of the harmonic oscillator, i.e., mixtures of Fock states with decreasing weight for increasing photon number \cite{dodonov}. 

\begin{figure}[h!]
	\includegraphics[trim = 0cm 0cm 0cm 0cm, clip, width=0.6\columnwidth]{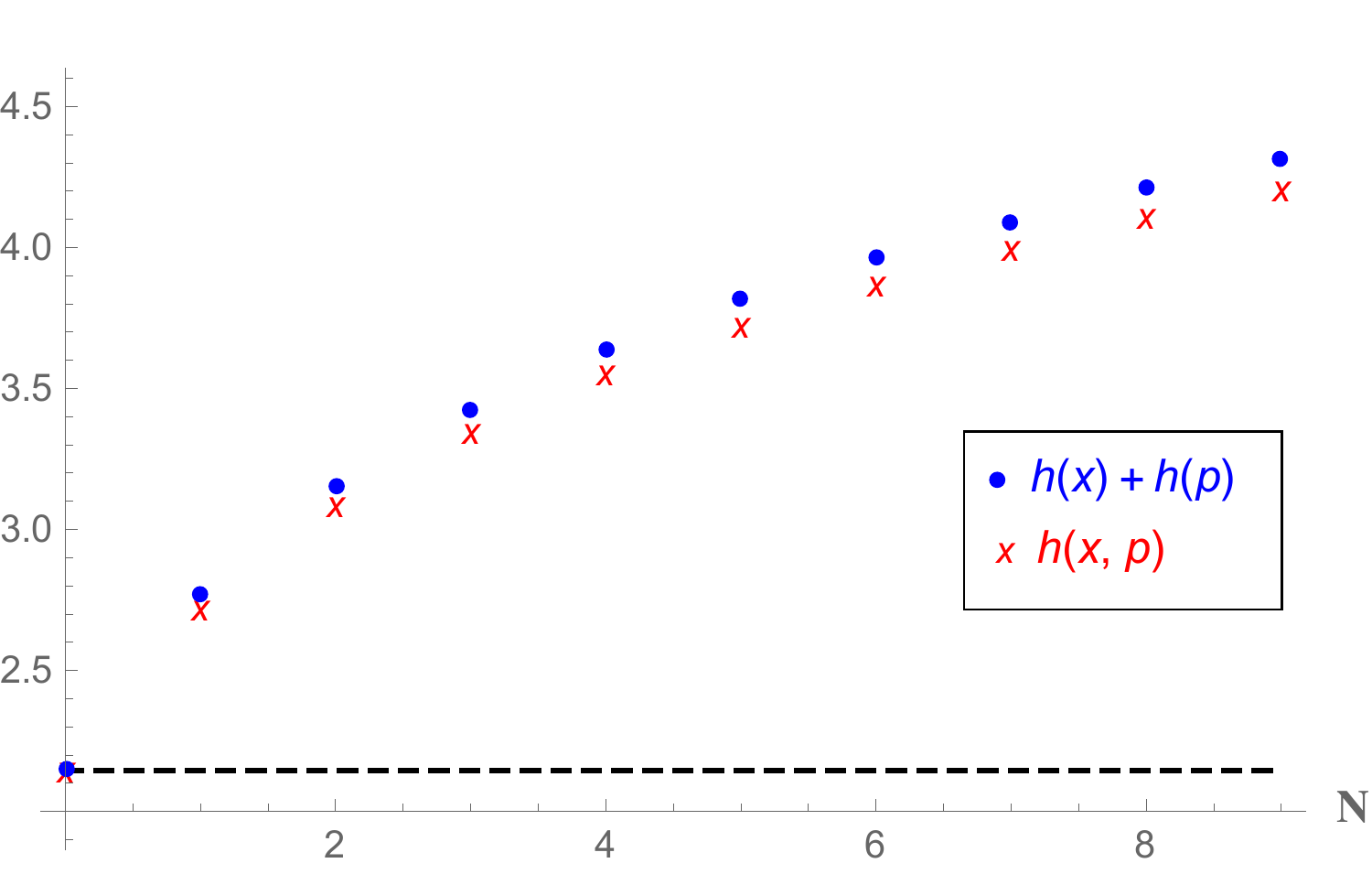}
	\caption{\label{passive} Test of the  uncertainty relation (\ref{conj1form1}) based on the joint entropy
		for extremal passive states, with $N$ being the highest photon number of the state. The blue dots correspond to $h(x)+h(p)$, the red dots correspond to $h(x,p)$, while the dashed line is the lower bound $\ln(\pi e)$ [we take $\hbar=1$]. }
\end{figure}

In Figure \ref{passive}, we consider \textit{extremal} passive states (i.e., passive states with equal weights up to a certain photon number $N$ and vanishing weights for larger photon numbers) and have plotted the joint entropy $h(x,p)$ as a function of $N$, see red dots. The dashed line is the lower bound $\ln(\pi e\hbar)$, so we clearly see that the uncertainty relation (\ref{conj1form1}) is obeyed. Since $h(x,p)$ is concave in the state, proving (\ref{conj1form1}) for extremal passive states would actually suffice to prove it for all passive states. For comparison with the regular entropic uncertainty relation (\ref{birula}), we have also plotted $h(x)+h(p)$, see blue dots, which illustrates that our rotation-invariant uncertainty relation provides an improvement. Although the improvement is minor in this example, it is worth noting that Eq. (\ref{conj1form1}) takes into account some $x$-$p$ correlations that are not visible in the second-order moments (all passive states have $\sigma_{xp}=0$), so no improvement at all would be obtained with our entropic uncertainty relation (\ref{conjecture}) relying on the covariance matrix.

We have also numerically tested other states with positive Wigner functions which are closer to the bound, such as mixtures of two squeezed states, and relation (\ref{conj1form1}) was verified in every tested case.

\subsection{Numerical tests of the  uncertainty relations (\ref{conjecture}) and (\ref{conjecture-bis})}

\begin{figure}[h!]
	\includegraphics[trim = 0cm 0cm 0cm 0cm, clip, width=0.58\columnwidth]{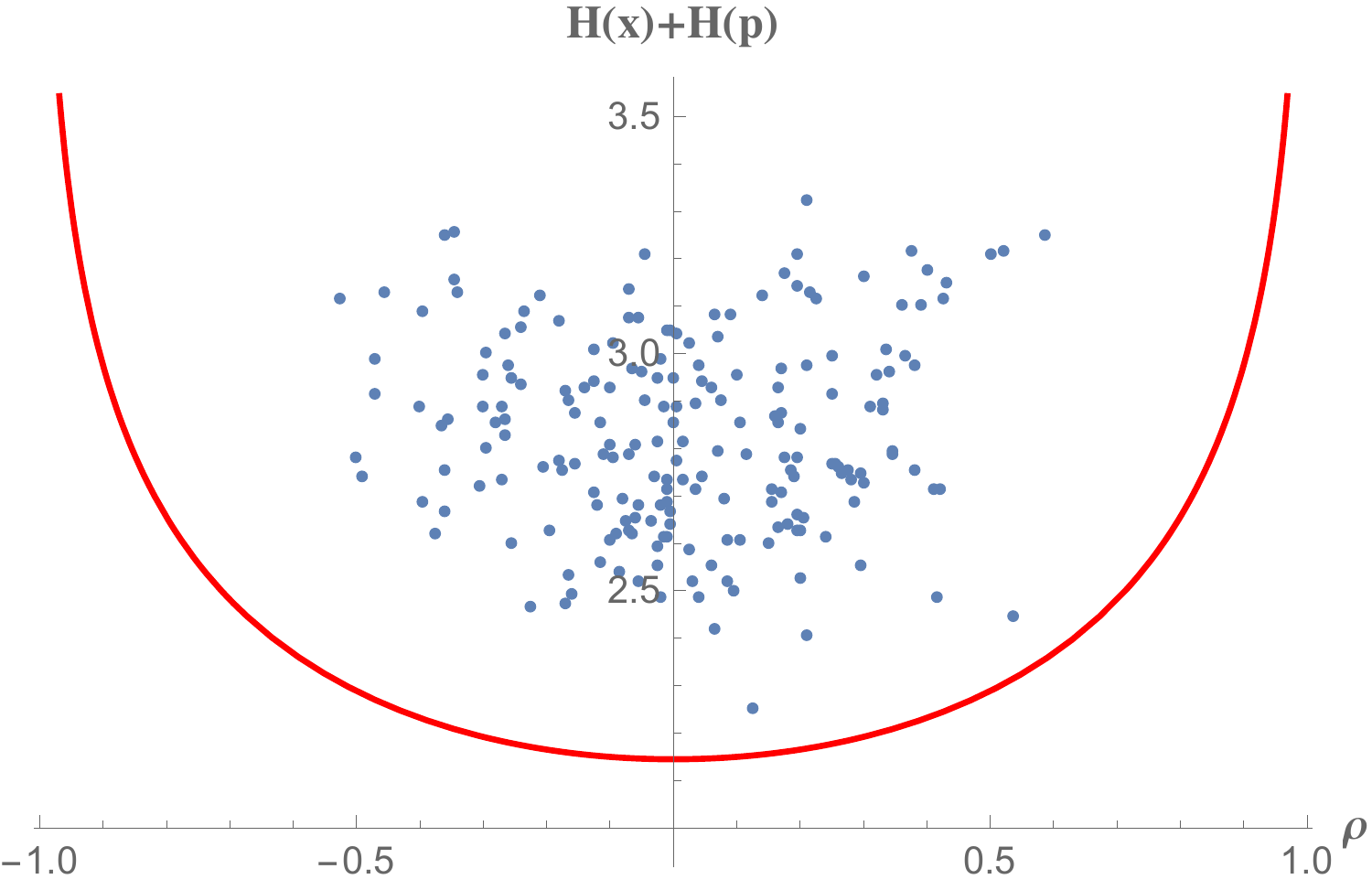}
	\caption{\label{testmatU} Test of the tight entropic uncertainty relation~(\ref{conjecture}) for random pure states generated by applying a $4\times 4$ random unitary onto the vacuum state. The blue dots correspond to $h(x)+h(p)$, while the red curve represents the improved lower bound $\ln(\pi e)+I_G(x{\rm :}p)$ [we take $\hbar=1$]. All quantities are plotted as a function of the correlation coefficient $\rho$.}
\end{figure}

We have also conducted many numerical tests in order to verify the accuracy of the tight entropic uncertainty relation. For numerical purposes, it was simpler to  consider the uncertainty relation in its form with differential entropies, eq.~(\ref{conjecture}). First, we have considered random pure states, which we generated by applying a random unitary transformation to the vacuum state. In Figure \ref{testmatU}, each blue dot corresponds to $h(x)+h(p)$ as computed for a random state generated with a $4\times 4$ unitary matrix (each state belongs to the space spanned by the Fock states $\ket{n}$ with $n=0,1,2,3$). The red curve represents the improved lower bound on $h(x)+h(p)$ that results from Eq. (\ref{conjecture}), namely
$\ln(\pi e\hbar)+I_G(x{\rm :}p)$. Here, the Gaussian mutual information is expressed as 
\begin{equation}
I_G(x:p)=-\frac{1}{2}\ln\left(1-\rho^2\right)
\end{equation}
where $\rho=\sigma_{xp}/(\sigma_x\sigma_p)$ stands for the correlation parameter. We clearly see that all points lie above the improved lower bound, corroborating the new entropic uncertainty relation (\ref{conjecture}). Note that other tests have been carried out with unitary transformations of greater dimensions, but this generally yields states with greater values of $h(x)+h(p)$, which are less interesting for verification purposes.

\begin{figure}[h!]
	\includegraphics[trim = 0cm 0cm 0cm 0cm, clip, width=0.58\columnwidth]{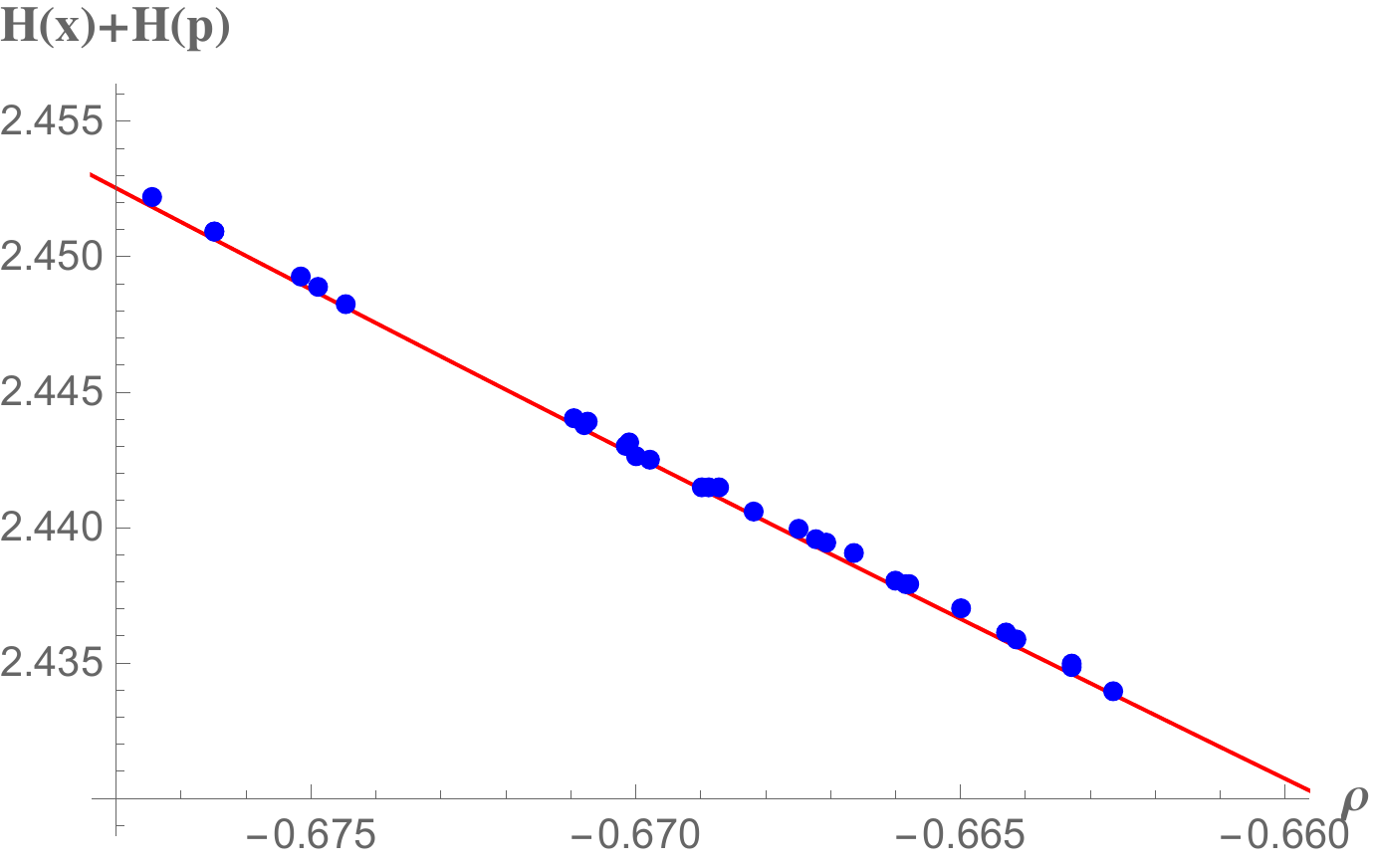}
	\caption{\label{slightlyNGstatesZoom} Test of the tight entropic uncertainty relation~(\ref{conjecture}) for slightly non-Gaussian states of the form $\ket{\psi}\propto (\ket{s}+\epsilon\ket{\phi})$ where $\ket{s}$ is a squeezed state (with $s=1.5$) along an axis rotated by an angle of $\theta=\pi/4$, $\ket{\phi}$ is a random pure state as in Fig. \ref{testmatU}, and $\epsilon=0.01$. The blue dots correspond to $h(x)+h(p)$, while the red curve represents the improved lower bound  $\ln(\pi e)+I_G(x{\rm :}p)$ [we take $\hbar=1$]. All quantities are plotted as a function of the correlation coefficient $\rho$. A zoom in of the interesting region is shown in this figure .}
\end{figure}

As a more stringent test, we have computed $h(x)+h(p)$ for some slightly non-Gaussian pure states lying in the neighborhood of the Gaussian pure states that saturate the uncertainty relation. To do so, we generated states of the form $\ket{\psi}\propto(\ket{s}+\epsilon\ket{\phi})$ where $\ket{s}$ is a squeezed state, $\ket{\phi}$ is any other pure state and $\epsilon \ll 1$. In Figure \ref{slightlyNGstatesZoom}, we have chosen $\ket{\phi}$ as some random pure state generated by the above method, $\epsilon=0.01$, and a squeezed state $\ket{s}$ along an axis rotated by an angle of $\theta=\pi/4$ with the $x$-axis (with a squeezing parameter $s\equiv e^r=1.5$). Its wave function has the form
\begin{equation}
\langle x | s \rangle =\sqrt[4]{\frac{2 s^2}{\pi  \left(s^4+1\right)}} \exp \left(\frac{i \left(s^2+i\right) x^2}{2 \left(s^2-i\right)}\right)
\end{equation}
which is non-Gaussian, implying that it cannot saturate the ordinary entropic uncertainty relation (\ref{birula}).
We have verified that, even if they lie very close to the boundary, all states $\ket{\psi}$ verify the tight entropic uncertainty relation. 
Similar simulations have also been performed with squeezed states of different parameters and with different values of $\epsilon$, yet no counterexample was found. 


\subsection{Concavity of the uncertainty functional}

The regular entropic uncertainty relation (\ref{birula}) was proven for pure states in \cite{Birula}. However, since the differential entropy is a concave function of the probability distribution, it is valid to mixed states as well (as mentioned in \cite{Birula}). Decomposing a mixed state into pure states, the concavity implies that pure states are the ``worst cases'', i.e., the lowest value of the functional $h(x)+h(p)$. Naturally, we also need to investigate the concavity of our new uncertainty functionals.  For our conjectured rotation-invariant uncertainty relation based on the joint entropy, we know that the joint differential entropy is concave since we limit ourselves to positive Wigner functions, which can be viewed as classical joint probability distributions. Hence, the left-hand side term of Eq. (\ref{conj1form1}) is a concave function of the state.

\begin{figure}[h!]
	\includegraphics[trim = 0cm 0cm 0cm 0cm, clip, width=0.7\columnwidth]{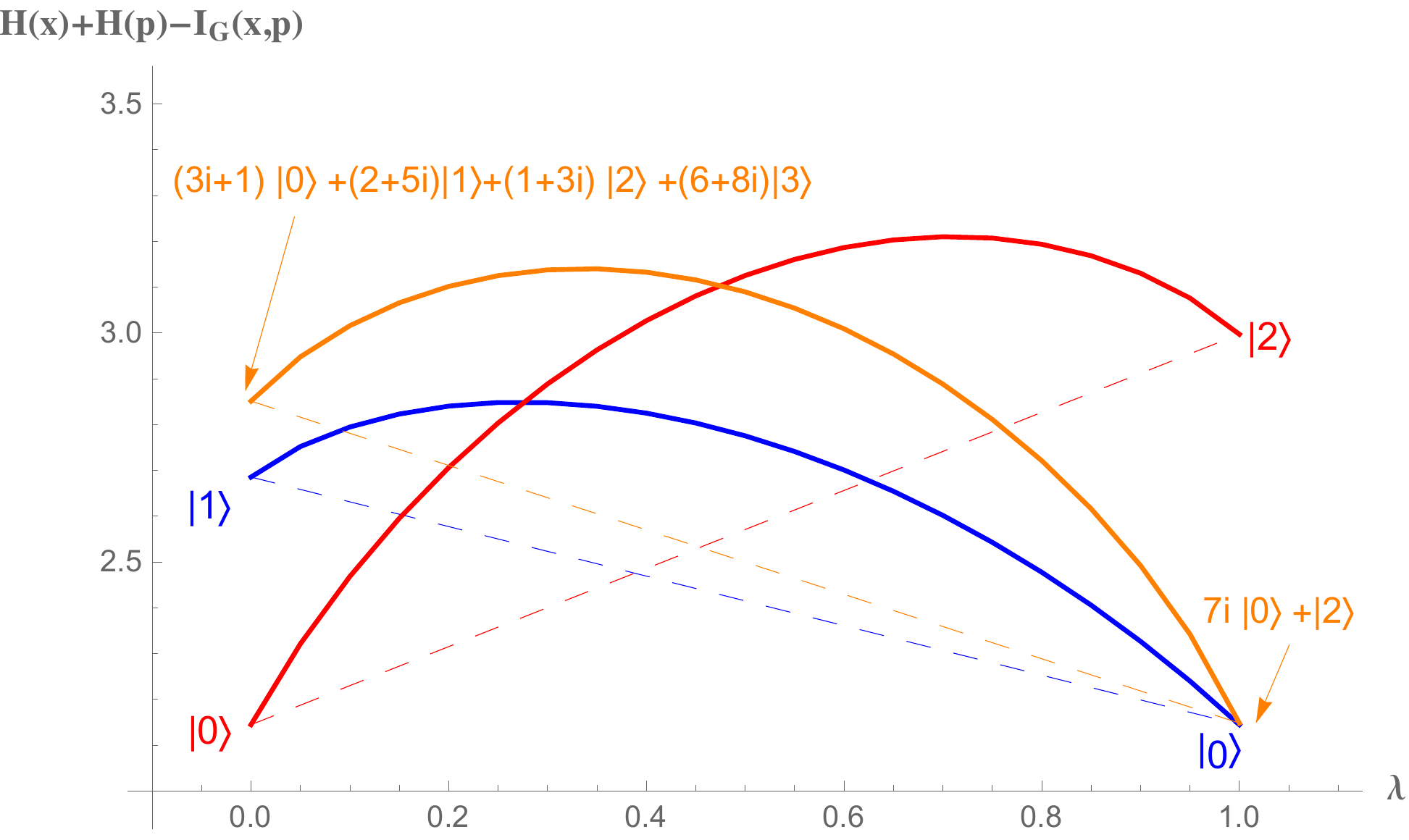}
	\caption{\label{concavity} Test of the concavity of the uncertainty functional $F(\rho)$ used in relation (\ref{conjecture}). We consider three different binary mixtures tuned by parameter $\lambda$: $\lambda \ket{0}\bra{0}+(1-\lambda) \ket{1}\bra{1}$, $\lambda \ket{2}\bra{2}+(1-\lambda) \ket{0}\bra{0}$, and $\lambda \ket{\psi}\bra{\psi}+(1-\lambda) \ket{\phi}\bra{\phi}$, where $\ket{\psi}=7i\ket{0}+\ket{2}$ and $\ket{\phi}=(3i+1)\ket{0}+(2+5i)\ket{1}+(1+3i)\ket{2}+(6+8i)\ket{3}$.}
\end{figure}

In contrast,  it seems hard to prove the concavity of the uncertainty functional $F(\hat \rho)$ of Eq. (\ref{uncertainty-functional}) which appears in the left-hand side of the tight entropic uncertainty relation (\ref{conjecture}). This is because while $h(x)$ and $h(p)$ are concave, $I_G(x{\rm :}p)$ is not convex. And even if it is known that $\log(|\gamma|)$ is concave \cite{Cover}, nothing can be said about $\log(\sigma_x^2\sigma_p^2)$. Nevertheless, numerical tests corroborate the fact  that $F(\hat \rho)$ is a concave function of the state. As an example,  we have analyzed mixtures of two pure states of the form $\lambda \ket{\psi_1}\bra{\psi_1}+(1-\lambda)\ket{\psi_2}\bra{\psi_2}$, with $0\le \lambda\le 1$. In Figure~\ref{concavity}, we have numerically verified that $F(\lambda \ket{\psi_1}\bra{\psi_1}+(1-\lambda)\ket{\psi_2}\bra{\psi_2})\geq\lambda F(\ket{\psi_1}\bra{\psi_1})+(1-\lambda)F(\ket{\psi_2}\bra{\psi_2})$.

Interestingly, we can prove the concavity of $F(\hat \rho)$ in some special case by using the expression of the entropic uncertainty relation in terms of 
non-Gaussianity measures based on relative entropies, Eq. (\ref{conjecture2}). We consider the mixture of two states that have the same first- and second-order moments. Hence, the right-hand side term of Eq. (\ref{conjecture2}) is constant and we need to prove that
\begin{eqnarray}
D(\lambda x_1+(1-\lambda)x_2\,||\,[\lambda x_1+(1-\lambda)x_2]_G)\leq\lambda  D(x_1\,||\,[x_1]_G)+(1-\lambda)D(x_2\,||\,[x_2]_G)
\end{eqnarray}
where $[x]_G$ means that we take the Gaussian distribution that leads to the same variance as the probability distribution of $x$. (Of course, we have an identical inequality for the $p$ quadrature.) By comparison, the convexity of the relative entropy implies that
\begin{eqnarray}
D(\lambda x_1+(1-\lambda)x_2\,||\,\lambda [x_1]_G+(1-\lambda)[x_2]_G) \leq\lambda  D(x_1\,||\,[x_1]_G)+(1-\lambda)D(x_2\,||\,[x_2]_G)
\end{eqnarray}
which is equivalent to the previous inequality since we mix up distributions with the same first- and second-order moments.

Remark that the uncertainty relation (\ref{conjecture}) is invariant under displacements, so that, with no loss of generality, we only need to consider states with zero mean values. Thus, we have proven the concavity of $F(\hat \rho)$ when two states with the same covariance matrix are mixed.
Yet, in the general case, we have not been able to prove the concavity.


\section*[appendix]{APPENDIX B : Partial proof of equation (\ref{nmodes}) }

 The proof follows the same variational method used in the one-mode case, that is, we prove that any $n$-mode squeezed vacuum state is a local minimum of the uncertainty functional 
\begin{equation}
F(\hat\rho) = h(\vec x)+h(\vec p)   -  {1\over 2} \ln \left(\frac{|\gamma_s||\gamma_p|}{ |\gamma| } \right)
\label{uncertainty-functional-n-mode}
\end{equation}
Since $F(\hat\rho)$ is invariant under $(\vec x,\vec p)$-displacements, it will imply that all Gaussian pure states are similarly local minima. Note that we assume, as for the one-mode case, that these are the unique solutions of our minimization problem and that the uncertainty functional $F(\hat\rho)$ is concave in $\hat \rho$, so that (\ref{nmodes}) is valid for mixed states as well.

We seek for an $n$-mode pure state $\ket{\psi}$ that minimizes the functional $F(\proj{\psi})$ with constraints on the  normalization of  $\ket{\psi}$ and mean values of $\vec x$ and $\vec p$ quadratures. We use the Lagrange multiplier method with
\begin{equation}
J=h(\vec x)+h(\vec p) -\frac{1}{2}\ln|\gamma_x|-\frac{1}{2}\ln|\gamma_p|+{1\over 2}\ln|\gamma| +\lambda( \mean{\psi|\psi}-1) +\sum_{i=1}^{2n}\mu_i \mean{\psi|\hat{r_i}|\psi}.
\end{equation}
Here, $\lambda$ and $\mu_i$  are Lagrange multipliers, while the elements of the covariance matrix $\gamma$ can be expressed as $\gamma_{ij}=~\mean{\psi|\hat{r}_i\hat{r}_j+~\hat{r}_j\hat{r}_i|\psi}/2$ since the states are normalized and centered on $0$. As in the one-mode case, 
we solve the variational  equation ${{\partial J}\over{\partial \bra{\psi}}}=0$, so we write the derivative of each term
\begin{equation}
\frac{\partial h(\vec{x})}{\partial \bra{\psi}}=-\left(  \ln W_x(\vec{x}) +1\right) \ket{\psi} \qquad\qquad \frac{\partial h(\vec{p})}{\partial \bra{\psi}}=-\left(  \ln W_p(\vec{p}) +1\right) \ket{\psi}.
\end{equation}
For the three terms involving the derivative of the determinant of a matrix, we use Jacobi's formula so that
\begin{eqnarray}
\frac{\partial }{\partial \bra{\psi}}\ln|\gamma_x|&=&\frac{1}{|\gamma_x|}   \frac{\partial }{\partial \bra{\psi}}|\gamma_x|
=\frac{1}{|\gamma_x|} \mathrm{Tr}\left[|\gamma_x|\gamma_x^{-1}\frac{\partial \gamma_x}{\partial \bra{\psi}}\right] 
=\sum_{i=1}^{n}\sum_{j=1}^{n}\gamma_{x_{ik}}^{-1}\frac{\partial\gamma_{x_{ki}}}{\partial\bra{\psi}}
=\sum_{i=1}^{n}\sum_{j=1}^{n}\gamma_{x_{ik}}^{-1}\frac{(\hat{x}_k\hat{x}_i+\hat{x}_i\hat{x}_k)}{2}\ket{\psi}\nonumber\\
&=&\left[\sum_{i=1}^{n}\sum_{j=1}^{n}\frac{\hat{x}_k \gamma_{x_{ik}}^{-1}\hat{x}_i}{2}+\sum_{i=1}^{n}\sum_{j=1}^{n}\frac{\hat{x}_i \gamma_{x_{ik}}^{-1}\hat{x}_k}{2} \right]\ket{\psi} 
=\vec{x}^T\gamma_x^{-1}\vec{x}\,\ket{\psi}.
\end{eqnarray}
where  we used the fact that $\gamma_{ik}^{-1}=\gamma_{ki}^{-1}$ since the matrix is symmetric.
Similary, we find
\begin{equation}
\frac{\partial }{\partial \bra{\psi}}\ln|\gamma_p|=\vec{p}^T\gamma_p^{-1}\vec{p}\,\ket{\psi}
\qquad\qquad
\frac{\partial }{\partial \bra{\psi}}\ln|\gamma|=\vec{r}^T\gamma^{-1}\vec{r}\,\ket{\psi}.
\end{equation}
Finally, the last terms give
\begin{eqnarray}
 \frac{\partial}{\partial \bra{\psi}} \bigg(  \lambda( \mean{\psi|\psi}-1) + \sum_{i=1}^{2n}\mu_i \mean{\psi|\hat{r}_i|\psi}  \bigg)    
= \left(\lambda+\sum_{i=1}^{2n}\mu_i \hat{r}_i\right)\ket{\psi}.
\end{eqnarray}
so that the variational equation can be rewritten as an eigenvalue equation for $ \ket{\psi}$,
\begin{equation}
 \bigg[-\ln W_x(\vec{x}) -\ln W_p(\vec{p}) -2+\lambda+\sum_{i=1}^{2n}\mu_i \hat{x}_i     
  - {1\over 2}\vec{x}^T\gamma_x^{-1}\vec{x}  - {1\over 2}\vec{p}^T\gamma_p^{-1}\vec{p} + {1\over 2}\vec{r}^T\gamma^{-1}\vec{r}
\bigg]  \ket{\psi}=0 . 
\label{eigenequ2}
\end{equation}

We now check that Eq. (\ref{eigenequ2}) is verified by $\ket{\psi}=\hat S \ket{0}$, that is, by any $n$-mode squeezed vacuum state. For such a state, the marginals of the Wigner functions are given by
\begin{eqnarray}
W_x(\vec{x})=((2\pi)^n|\gamma_x|)^{-{1\over 2}} \, e^{-  {1\over 2}\vec{x}^T\gamma_x^{-1}\vec{x}},  \qquad\qquad
W_p(\vec{p})= ((2\pi)^n|\gamma_p|)^{-{1\over 2}} \, e^{-  {1\over 2}\vec{p}^T\gamma_p^{-1}\vec{p}}, 
\end{eqnarray}
so that
\begin{equation}
\ln W_x(\vec{x}) + \ln W_p(\vec{p}) =-\ln\left((2\pi)^n \sqrt{|\gamma_x||\gamma_p|} \right)-  {1\over 2}\vec{x}^T\gamma_x^{-1}\vec{x}-  {1\over 2}\vec{p}^T\gamma_p^{-1}\vec{p}
\label{sumW}
\end{equation}
We apply  ${1\over 2} \,\vec{r}^T\, \gamma^{-1} \, \vec{r}$  on the squeezed vacuum state $\ket{\psi}$ by using the canonical transformation of $\vec r$ in the Heisenberg picture, namely
${\hat S}^{\dagger} \vec r\,{\hat S} =M \vec r$, we find
\begin{eqnarray}
{1\over 2} \,\vec{r}^T\, \gamma^{-1} \, \vec{r} \, \ket{\psi}&=&{1\over 2} \,\vec{r}^T\, \gamma^{-1} \, \vec{r} \, \hat S \ket{0}=
\frac{1}{2}{\hat S} \,\vec r^T \,M^T\gamma^{-1}M\vec r\,\ket{0}
=\frac{1}{2}{\hat S}   \,\vec r^T\,\gamma_\mathrm{vac}^{-1} \vec r \,  \ket{0}  =
{\hat S}\ket{0} = \ket{\psi}
\end{eqnarray}
since the covariance matrix $\gamma$ of state $\ket{\psi}$ can be expressed as $\gamma = M \gamma_\mathrm{vac} M^T$. This implies that state $\ket{\psi}$ is an eigenvector of ${1\over 2} \,\vec{r}^T\, \gamma^{-1} \, \vec{r}$  with eigenvalue 1. Therefore, using this result together with equation (\ref{sumW}),  the eigenvalue equation for $\ket{\psi}$ can be written as 
\begin{eqnarray}
\left[\ln\left((2\pi)^n \sqrt{|\gamma_x||\gamma_p|} \right)-1+\lambda+\sum_{i=1}^{2n}\mu_i\hat{r}_i\right]\ket{\psi}=0.
\end{eqnarray}
The value of $\lambda$ is found by multiplying this equation on the left by $\bra{\psi}$ and by using the constraints  $ \mean{\psi|\psi}=1$ and $\mean{\psi|\hat{r}_i|\psi}=0$ for all $i$, namely
\begin{eqnarray}
\bra{\psi}  \left[\ln\left((2\pi)^n \sqrt{|\gamma_x||\gamma_p|} \right)-1+\lambda+\sum_{i=1}^{2n}\mu_i\hat{r}_i\right]\ket{\psi}
 =0\qquad \Rightarrow\qquad \lambda=1-\ln\left((2\pi)^n \sqrt{|\gamma_x||\gamma_p|} \right).
\end{eqnarray}
We are now left with equation
\begin{equation}
\left[\sum_{i=1}^{2n}\mu_i\hat{r}_i\right]\ket{\psi}=0
\end{equation}
which is satisfied if we set all the $\mu_i=0$.

In conclusion, we have proven that, with the appropriate choice of  $\lambda$ and $\mu_i$, the $n$-mode squeezed vacuum states are solutions of Eq. (\ref{eigenequ2}), so they minimize our uncertainty functional $F(\proj{\psi})$. Since $F(\proj{\psi})$ is invariant under displacements, the displaced squeezed vacuum states are also solutions, so this minimization result encapsulates all pure Gaussian states. We  find the minimum value $n\ln(\pi e \hbar)$ by evaluating $F$ for any of these states.


\begin{thebibliography}{99}
	
	
	
	\bibitem{Heisenberg}  Heisenberg W 1927 Z. Phys. {\bf 43} 172 

	\bibitem{Kennard}  Kennard E H 1927 Z. Phys. {\bf44} 326


	\bibitem{Schrodinger}  Schr\"{o}dinger E 1930	Preuss. Akad. Wiss. {\bf 14} 296
	\bibitem{Robertson}  Robertson H P 1930 Phys. Rev. {\bf 35} 667A
	


	\bibitem{Birula} Bialynicki-Birula I and Mycielski J 1975 Commun. Math. Phys. {\bf 44} 129 
	
	\bibitem{Deutsch}  Deutsch D 1983 Phys. Rev. Lett. {\bf 50} 631 
	\bibitem{Kraus}  Kraus K 1987  Phys. Rev. D {\bf 35} 3070 
	\bibitem{MaassenUffink} Maassen H and  Uffink J B M 1988 Phys. Rev. Lett. {\bf 60} 1103
	
	\bibitem{Birula2}  Bialynicki-Birula I and Rudnick L 2011 {\it Statistical Complexity} ed. K D Sen (Springer, Berlin) pp. 1-34 

\bibitem{Coles2} Coles P J, Berta M, Tomamichel M and Wehner S 2017 Rev. Mod. Phys. {\bf89}(1) 15002
		
	\bibitem{Renes} Renes J M and  Boileau J-C 2009  Phys. Rev. Lett. {\bf 103} 020402
	\bibitem{Berta}	Berta M, Christandl M, Colbeck R,	Renes J M and Renner R 2010 Nature Phys. {\bf 6} 659 
	
	\bibitem{Tomamichel1} Tomamichel M and Renner R 2011 Phys. Rev. Lett. {\bf 106}110506 
	\bibitem{Tomamichel2} Tomamichel M, Lim C C W, Gisin N and Renner R 2010 Nature Commun. {\bf 3} 634 
		
	\bibitem{GrosshansCerf} Grosshans F and Cerf N J 2004 Phys. Rev. Lett. {\bf 92} 047905
	\bibitem{Furrer} Furrer F,  Franz T, Berta M,  Leverrier A,  Scholz V B,  Tomamichel M and  Werner R F 2012 Phys. Rev. Lett. {\bf 109} 100502
	
	
		
	\bibitem{duan}L. Duan M, Giedke G, Cirac J I and  Zoller P 2000 
	Phys. Rev. Lett. {\bf 84} 2722
	
	\bibitem{simon} Simon R 2000
	Phys. Rev. Lett. {\bf 84} 2726
	
	\bibitem{walborn}  Walborn S P, Taketani B G, Salles A,  Toscano F and  de Matos Filho R L 2009
	Phys. Rev. Lett. {\bf 103} 160505
	
	\bibitem{Coles}  Coles P J, Kaniewski J and  Wehner S 2014 Nature Communications {\bf 5} 5814
	
	


\bibitem{Genoni} Genoni M G, Paris M G A and Banaszek K  2008 Phys. Rev. A {\bf 78} 060303(R) 


	
\bibitem{Son} Son W 2015
Phys. Rev. A {\bf 92} 012114 

\bibitem{Cover} Cover T M and Thomas J A 2006
{\it Elements of Information Theory} New York Wiley 

	
\bibitem{Brocker} Br\"ocker T and  Werner R F 1995 J. Math. Phys. {\bf 36} 62 	

\bibitem{Hall} Hall M J W 1994 Phys. Rev. A {\bf 49} 42 

\bibitem{huang} Huang Y 2011 Phys. Rev. A {\bf 83} 052124

\bibitem{dodonov} Dodonov V V 2002 J. Opt. B {\bf 4} S98


	\bibitem{hertz} Hertz A,  Karpov E, Mandilara A and  Cerf N J 2016 Phys. Rev. A {\bf 93} 032330 
	\bibitem{katerina} Mandilara  A and Cerf N J 2012 Phys. Rev. A {\bf 86} 030102(R)

	\bibitem{Jizba}  Jizba P, Ma Y,  Hayes A and  Dunningham J A 2016 Phys. Rev. E {\bf 93} 060104 
	
\end{thebibliography}
\end{document}